\documentclass[10pt,twocolumn]{article}
\usepackage[utf8]{inputenc}
\usepackage{fancyhdr}

\usepackage{lmodern}			
\usepackage[T1]{fontenc}		
\usepackage{indentfirst}		
\usepackage{color}			
\usepackage{graphicx}			
\usepackage{microtype} 			
\usepackage[portuguese]{babel}
\usepackage{amsbsy}			
\usepackage{amsfonts}			
\usepackage{amsmath}			
\usepackage{amssymb}			
\usepackage{amstext}			
\usepackage{model}
\usepackage{times}
\usepackage{epsfig}
\usepackage{graphicx}
\usepackage{amsmath}
\usepackage{amssymb}
\usepackage{color}
\usepackage{comment}
\usepackage{colortbl}
\usepackage{multirow}
\usepackage{array}
\usepackage[dvipsnames,table]{xcolor}
\usepackage{pdflscape}
\usepackage{comment}
\usepackage[final]{pdfpages}
\usepackage{afterpage}
\definecolor{bblue}{rgb}{0.0, 0.3, 0.9}

\definecolor{ggray}{rgb}{0.8, 0.8, 0.8}
\usepackage[pagebackref=true,breaklinks=true,colorlinks,bookmarks=false]{hyperref}

\newcommand{\ABACODP}[1]{%
\thicklines
\begin{picture}(8,0)
    \ifcase#1{   
       \put(0,0)    {\line(1,0){4}}
       \multiput(5,0)(2,0){2}{\oval(2,4)}}
    \or{         
       \put(2,0)    {\line(1,0){4}}
       \multiput(1,0)(6,0){2}{\oval(2,4)}}
    \fi
\end{picture}
    } 

\newcommand{\ABACODG}[1]{%
\thicklines
\begin{picture}(14,0)
    \ifcase#1{   
       \multiput(1,0)(2,0){5}{\oval(2,4)}}
       \put(10,0)   {\line(1,0){4}}
    \or{         
       \multiput(1,0)(2,0){4}{\oval(2,4)}}
       \put(8,0)   {\line(1,0){4}}
       \put(13,0)   {\oval(2,4)}
    \or{         
       \multiput(1,0)(2,0){3}{\oval(2,4)}
       \put(6,0)   {\line(1,0){4}}
       \multiput(11,0)(2,0){2}{\oval(2,4)}}
    \or{         
       \multiput(1,0)(2,0){2}{\oval(2,4)}
       \put(4,0)   {\line(1,0){4}}
       \multiput(9,0)(2,0){3}{\oval(2,4)}}
    \or{         
       \put(1,0)  {\oval(2,4)}}
       \put(2,0)   {\line(1,0){4}}
       \multiput(7,0)(2,0){4}{\oval(2,4)}
    \fi
\end{picture}
    } 
       
\newcommand{\ABACOD}[1]{%
    \ifnum#1>9
       \errmessage{#1: Argumento invalido para ABACO}
    \fi
    \ifnum#1<0
       \errmessage{#1: Argumento invalido para ABACO}
    \fi
\begin{picture}(24,0)
    \ifnum#1<5
       \put(16,0) {\ABACODP{0}}
    \else   
       \put(16,0) {\ABACODP{1}}
    \fi
    \ifnum#1<5
       \put(0,0)  {\ABACODG{#1}}
    \else
       \ifcase#1\or \or \or \or
          \or  \put(0,0)  {\ABACODG{0}}
          \or  \put(0,0)  {\ABACODG{1}}
          \or  \put(0,0)  {\ABACODG{2}}
          \or  \put(0,0)  {\ABACODG{3}}
          \or  \put(0,0)  {\ABACODG{4}}
       \fi
    \fi   
\end{picture}
    } 
    
\cvprfinalcopy 

\ifcvprfinal\fi

\begin{document}

\title{Agrupamento de manchetes - Cluster}
\author{Ciro Javier Diaz Penedo\thanks{Do Instituto de Matemática, Estatística e Computação Científica da Universidade de Campinas (Unicamp). \textbf{Contato}: \tt\small{ra153868@ime.unicamp.br  e  ra153866@ime.unicamp.br}}  \\
Lucas Leonardo Silveira Costa$^{*}$  }

\maketitle
\begin{abstract}
Neste trabalho tratamos do problema de agrupamento em \textit{headlines} (manchetes) do jornal ABC (\textit{Australian Broadcasting Corporation}) utilizando técnicas de aprendizado de máquina sem supervisão. Apresentamos e discutimos os resultados sobre os \textit{clusters} encontrados.
\end{abstract}

\section{Introdução}

Atualmente o processo de detecção de padrões é frequentemente necessário em empresas, órgãos governamentais, bancos, etc. Tais processos são úteis para realizar políticas de decisões, por exemplo, um banco pode detectar padrões em seus clientes e definir taxas de serviços, políticos podem avaliar as características de seus candidatos e dos não-candidatos e assim investir em campanhas para conquistar mais eleitores, entre outras. 

Esse processo para detectar padrões pode ser realizado por meio de aprendizado de máquina sem supervisão (\textit{unsupervised learning}) utilizando agrupamento (\textit{clustering}), isto é, não se conhece as características dos grupos existentes e nem a quantidade de elementos em cada grupo.

Uma aplicação útil usando agrupamento é a detecção de padrões em textos. O agrupamento em textos pode ser útil na pesquisa forense atuando na detecção de mensagens de criminosos \cite{re3} e na detecção de spam na caixa de \textit{e-mail}.

Nesse trabalho utilizamos o conceito de \textit{N-grams} \cite{re1} para a extração de atributos em textos e realizamos a detecção de padrões em \textit{headlines} (manchetes) do jornal ABC (\textit{Australian Broadcasting Corporation} - \url{http://www.abc.net.au/}), depois utilizamos o método \textit{K-means} para agrupar as manchetes. Para aplicar este método é necessário transformar cada texto em um vetor numérico de atributos (\textit{features}) que representam características relevantes deles para relacionar um texto (vetor) com outro.

A metodologia escolhida para encontrar a solução do problema é descrita a seguir:

\vspace{1mm}

\noindent \ \ i. Simplificar a linguagem natural das \textit{headlines};

\noindent \ ii. Construir um dicionário de \textit{N-grams};

\noindent iii. Construir os vetores de atributos para cada \textit{headline};

\noindent \ iv. Aplicar o algoritmo de agrupamento (\textit{k-means} no caso);

\noindent \ \ v. Discutir os resultados dos experimentos.





\section{Simplificar a linguagem natural} \label{Sec2}

  A linguagem natural tem muitas regras para melhorar o entendimento entre as pessoas. Como o nosso objetivo é relacionar textos, precisamos simplifica-los ficando com apenas o essencial. Por isso todas as palavras vão ser consideradas em minúsculas, tiramos todos os símbolos que não estejam no alfabeto e mantemos apenas o radical \cite{re4} dos substantivos, adjetivos, verbos e advérbios. Assim o texto:
  \textit{"João gosta de assistir filmes. Maria gosta de filmes também."} torna-se \textit{"joão gost assist film maria gost film também"} 

\section{Construir um dicionário de \textit{N-grams}}
\label{Sec3}
Um \textit{N-gram} \cite{re1} é um conjunto ordenado de $N$ palavras e um dicionário destes seria uma sequência de tuplas formadas por uma chave e uma descrição, em nosso a chave (\textit{key}) é um \textit{N-gram} e a descrição é a quantidade de vezes (\textit{frequência}) que este aparece no conjunto de dados. Para descrever a construção do dicionário suponhamos que temos estas duas frases no conjunto de dados:

(1) João gosta de assistir filmes. Maria gosta de filmes também."

(2) João também gosta de assistir a jogos de futebol !"

Após simplificar os dados via Seção \ref{Sec2} o nosso dicionário de \textit{1-grams} seria: 

\textit{ ``joão'':2, ``gost'':3, `` assist'':2, "film":2, "maria":1, "também":1, "jog":1, "futebol": 1 }. 

Para \textit{2-grams} teríamos: 

\textit{ ("joão","gost"):1, ("gost","assist"):2, ("assist",film"):1 ... }
\section{Construir os vetores de atributos.}
\label{Sec4}
Nos experimentos computacionais usamos os \textit{N-grams} do dicionário que possuem uma frequência maior ou igual a 10, mas para o exemplo anterior usaremos todas.

\begin{table}[ht]
\centering
\scalebox{1}{
\begin{tabular}{|l|c|c|c|c|c|c|c|}
\hline
\rowcolor{cyan}
Posição & 1ª & 2ª & 3ª & 4ª \\
\hline
Palavra & joão & gost & assist & film \\
\rowcolor{cyan}
Posição & 5ª & 6ª & 7ª & 8ª \\
\hline
Palavra &  maria & também & jog & futebol\\

\hline
\end{tabular}}
\caption{Tabelas com as palavras do dicionário} \label{1}
\end{table}

Esses oito \textit{1-grams} da Tabela \ref{1} representam os nossos atributos utilizados para a conversão dos textos em vetores, sendo assim, percorremos cada manchete e criamos um vetor binário de $8$ entradas que indica se o texto possui o \textit{1-gram} do dicionário ou não. Para o exemplo apresentamos o resultado na Tabela \ref{2}.

\begin{table}[ht]
\centering
\scalebox{1}{
\begin{tabular}{|l|c|c|c|c|c|c|c|c|c|c|c|}
\hline
\rowcolor{cyan}
Palavra & 1ª & 2ª & 3ª  & 4ª & 5ª & 6ª & 7ª & 8ª\\
\hline
\rowcolor{ggray}
Frase (1) & 1 & 1 & 1 & 1 & 1 & 1 & 0 & 0 \\ 
\hline
Frase (2) & 1 & 1 & 1 & 0 & 0 & 1 & 1 & 1 \\
\hline
\end{tabular}}
\caption{Amostras para realizar o cluster} \label{2}
\end{table}

Poderíamos também  usar \textit{2-gram} e nesse caso utilizaríamos o dicionário de \textit{2-grams}. Também podemos usar \textit{1-gram} + \textit{ 2-gram} juntando ambos dicionários. Claramente que \textit{N-grams} de $3$, $4$ ou mais palavras podem ser usados (e seria desejável), neste trabalho por problemas de potência computacional decidimos ficar com \textit{1-grams} e \textit{2-grams}.

\section{Agrupamento (\textit{K-means})}
\label{Sec5}

O algoritmo \textit{K-means} pode ser descrito como um problema de otimização (\ref{eq1}).
\begin{equation} \label{eq1}
\mbox{min: } J(c^{1},\dots ,c^{m}, \mu_{1},\dots,\mu_{k}) = \sum_{i=1}^{m} dist(x^{(i)}, \mu_{c^{i}})
\end{equation}

\noindent em que a função $J$ é chamada de inércia \cite{re5}, $x^{(i)}$ são os dados, $c^{i} = 1, \dots, k$ são os índices que indicam qual o grupo que $x^{(i)}$ pertence, $\mu_{k}$ é o centróide do grupo $k$ e a \textit{dist} é alguma medida ,por exemplo: $|| \cdot ||_{1}$, $|| \cdot ||_{2}$, \textit{cosine distance}.

O algoritmo se chama \textit{K-means} pois a palavra \textit{means} significa médias e indica a maneira como $\mu_{k}$ foi calculado, isto é, para obter os centróides usamos a equação \ref{eq2} 

\begin{equation} \label{eq2}
\mu_{kj} = \dfrac{1}{t} \sum_{i=1}^{t} x^{(c^{i})}_{j}, \ \ j=1,\dots, M,
\end{equation}

\noindent  em que $M$ é a dimensão de $x^{(i)}$, e desse modo, cada entrada $j$ de $\mu_{kj}$ é a média das entradas $j's$ dos vetores $x^{(i)}$ pertencente o grupo $k$.

O algoritmo \textit{K-means} é descrito a seguir:

Defina os $k$ centróides iniciais e repita $i=1,\dots,$it. max.:
\begin{enumerate}
\item Atribua os índices $c^{i}$ em $x^{(i)}$ em que \\ $c^{i}$ = arg(min $dist(x^{(i)}, \mu_{c^{i}})$);
\item Atualize os centróides $\mu_{k}$ calculando a média dos $x^{(i)}$  pertencentes ao cluster $k$.
\end{enumerate}

Uma variação do algoritmo \textit{K-means} é o algoritmo \textit{k-medoids} e nesse caso, $\mu_{k}$ é o elemento $x^{(h)}$ do grupo $k$ o qual possui o menor valor da somatória entre as distancias de $x^{(h)}$ com os elementos do grupo $k$. Outra variação é o \textit{K-median} que utiliza $\mu_{k}$ como sendo a mediana do grupo $k$ e o \textit{Mini Batch K-means} é uma variação que utiliza parcelas (\textit{Batch}) dos danos para realizar os clusters, sendo bem útil quando se tem muitos dados e pouca potência computacional.

Para a escolha da quantidade de \textit{cluster} ($k$) utiliza-se a \textit{elbow rule} (regra do cotovelo). Podemos ver na equação \ref{1} que se consideramos um único \textit{cluster} o valor do minimo que a função alcançaria seria o maior valor possível, e em contrapartida se considerarmos cada ponto sendo um cluster o valor atingido será zero, pois o centróide do \textit{cluster} $k$ seria o próprio ponto. 

Desse modo, a \textit{elbow rule} consiste relacionar a quantidade de \textit{clusters} $k$ com o valor da inércia $J(c^{1},\dots ,c^{m}, \mu_{1},\dots,\mu_{k})$ em \ref{eq1}, e escolher o $k$ tal que a variação de $J$ de $k$ para $k+1$ seja pequena.

Observação: Outras regras usadas para a escolha da quantidade de \textit{cluster} são o \textit{Silhoutte} e \textit{Calinski Harabasz} \cite{re5} mas, não vamos usar elas neste trabalho.
\section{Experimentos e Discussões}

Os experimentos computacionais foram realizados em \textit{python} e o conjunto de dados era composto por 1 milhão e uma manchetes publicadas entre 19/02/2003 até 31/12/2017 pela ABC. Após simplificar a linguagem natural (\ref{Sec2}) e construir os dicionários de \textit{1-grams} e \textit{2-grams} (\ref{Sec3}), construímos nosso vetor de atributos (\ref{Sec4}) usando os \textit{N-grams} com frequências maiores ou iguais a $10$. Isso seria em torno de $25000$ features para cada dicionário. Aplicamos o algoritmo \textit{k-means} variando o parâmetro $k$ entre $2$ e $21$ para tentar aplicar a \textit{elbow rule} (\ref{Sec5}). Os experimentos foram separados da seguinte maneira: No primeiro usamos as features obtidas via \textit{1-grams} e no segundo juntamos os features obtidas via \textit{1-grams} + \textit{2-grams}. Após selecionar o melhor $k$ (Experimentos $1$ e $2$) fizemos um terceiro experimento onde dividimos o conjunto de \textit{headlines} por anos ($15$ subconjuntos no total), e aplicamos k-means em cada um deles.

\subsection{Features baseadas em 1-grams}
\label{sub61}
\begin{figure}[ht]
\hspace{-1cm}
\includegraphics[scale=0.2]{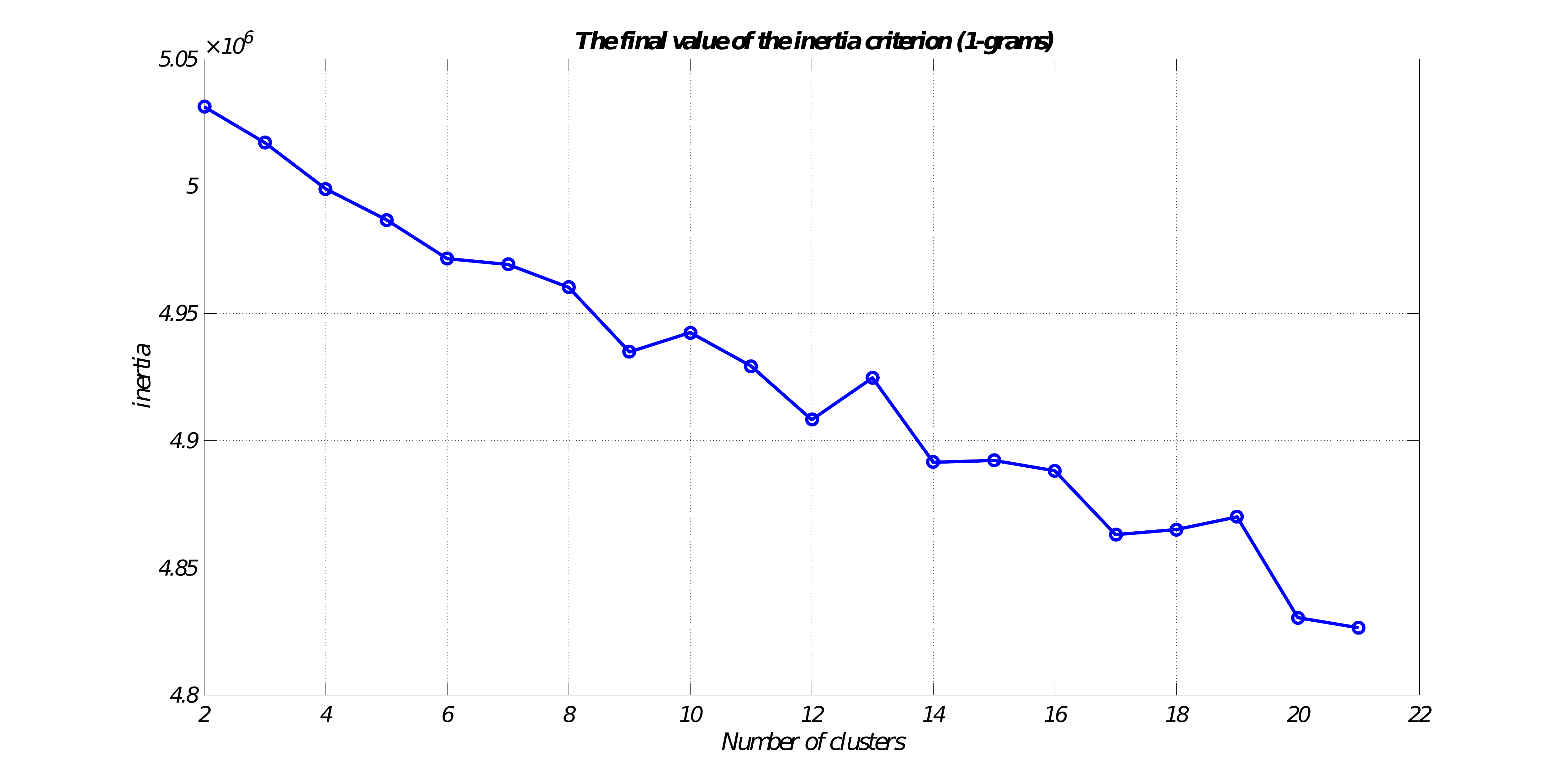}
\caption{Valores da função inércia \ref{eq1} em relação a quantidade de cluster considerada ($k=2,\dots,21$) considerando 1-\textit{gram}. \label{fig1}}   
\end{figure}

Na Figura \ref{fig1} podemos analisar o gráfico da função inércia (equação \ref{eq1}) para o caso 1-\textit{gram}. Nela observamos que de fato o valor da inércia vai diminuindo, porém, não é possível detectar se até $k=21$ o valor irá se estabilizar.


Na Tabela \ref{Tab1} apresentamos a quantidade de elementos em cada \textit{cluster} para $k = 20$ e $21$ do caso 1-\textit{gram}. Podemos observar que o \textit{cluster} 1 possui perto de $74\%$ dos dados, o resto dos dados se distribuem nos outros \textit{clusters}. Esse fato foi observado também em valores de $k$ menores que $20$. 

\begin{table}[!h]
\hspace{-1.6cm}
\scalebox{0.64}{ 
\begin{tabular}{|l|r|r|r|r|r|r|r|r|r|r|r|r|r|r|r|r|r|r|} 
\hline
\rowcolor{cyan}
Cluster & 1 & 2 & 3 & 4 & 5 & 6 & 7 & 8 & 9 & 10 & 11 \\
\hline
\rowcolor{ggray}
k = 20 & 763984 & 28373 & 24319 & 23418 & 20632 & 14089 & 14027 & 13973 & 12508 & 10865 & 10624 \\
\hline
k = 21 & 745966 & 37114 & 29213 & 27747 & 20618 & 18298 & 14718 & 11348 & 11282 & 10312 &  9542 \\
\hline
\hline
\rowcolor{cyan}
Cluster & 12 & 13 & 14 & 15 & 16 & 17 & 18 & 19 & 20 & 21 & - \\
\hline
\rowcolor{ggray}
k = 20 &  10458 &  9149 &  9139 &  9092 &  8772 &  5702 &  5598 &  4068 &  1211 &     -- &    -- \\
\hline
k = 21 &    8554 &  7836 &  7588 &  7109 &  6666 &  6193 &  6150 &  6097 &  5957 &  1693 &    --\\
\hline
\end{tabular}}
\caption{Quantidade de elementos em cada cluster, considerando $k = 20$ e $21$ e com atributos obtidos com 1-\textit{gram}.} \label{Tab1}
\end{table}

\subsection{Features com 1-grams + 2-grams}

\begin{figure}[ht]
\hspace{-0cm}
\includegraphics[trim={0 0 0 0}, scale=0.2]{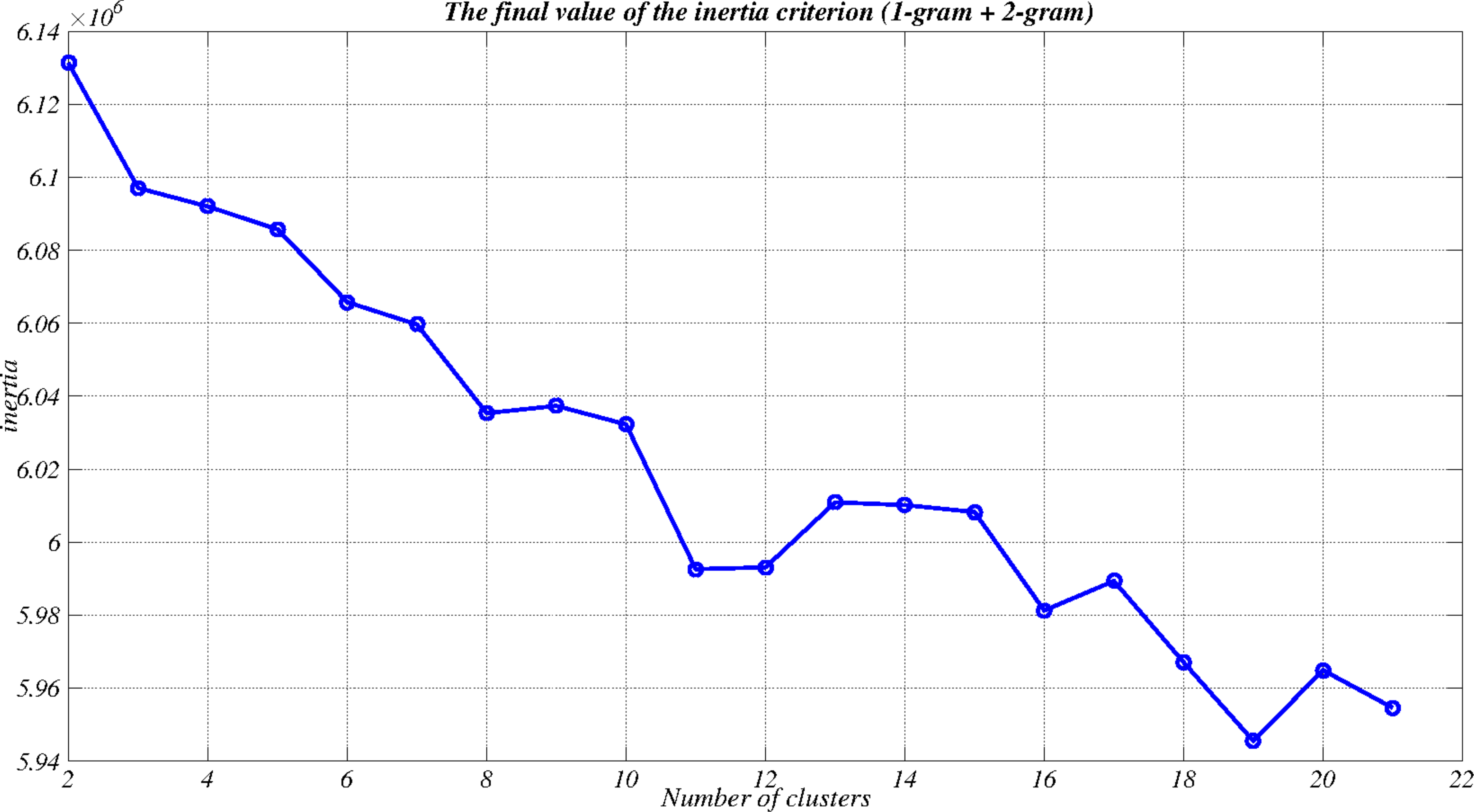}
\caption{Valores da função inércia \ref{eq1} em relação a quantidade de cluster considerada ($k=2,\dots,21$) considerando 1-\textit{gram} + 2-\textit{gram}. \label{fig2}}
\end{figure}

Na Figura \ref{fig2} apresentamos o gráfico da função inércia considerando os atributos extraídos com 1-\textit{gram} + 2-\textit{gram}. Como na Figura \ref{fig1} observamos que de fato o valor da inércia vai diminuindo, porém, não é possível detectar se até $k=21$ o valor irá se estabilizar.

Na Tabela \ref{Tab2} apresentamos a quantidade de elementos em cada clusters $k=18$ e $19$ para o caso este caso e vemos  que perto de $78\%$ dos dados estão localizados no primeiro \textit{cluster} como no Experimento descrito em \ref{sub61}.

\begin{table}[ht]
\hspace{-0.3cm}
\scalebox{0.64}{ 
\begin{tabular}{|l|r|r|r|r|r|r|r|r|r|r|r|r|r|r|r|r|r|r|} 
\hline
\rowcolor{cyan}
Cluster & 1 & 2 & 3 & 4 & 5 & 6 & 7 & 8 & 9 & 10  \\
\hline
\rowcolor{ggray}
k = 18 & 821361 & 32030 & 29574 & 19787 & 14207 & 11697 & 11501 & 10351 &  9611 &  8754 \\
\hline
k = 19 & 783575 & 29295 & 25300 & 21402 & 18469 & 18032 & 15779 & 14158 & 14002 & 11201 \\
\hline
\hline
\rowcolor{cyan}
Cluster & 11 & 12 & 13 & 14 & 15 & 16 & 17 & 18 & 19 & --  \\
\hline
\rowcolor{ggray}
k = 18   & 6387 &  6247 &  6212 &  4533 &  3425 &  3342 &   602 &   380 &   -- & -- \\
\hline 
k = 19  &  8790 &  8235 &  7265 &  7174 &  5924 &  4547 &  3444 &  2785 &   624 & --  \\
\hline
\end{tabular}}
\caption{Quantidade de elementos em cada cluster, considerando $k = 20$ e $21$ e com atributos obtidos com 1-\textit{gram} + 2-\textit{grams}.} \label{Tab2}
\end{table}

\subsection{Nuvem de palavras com 1-gram + 2-grams}

\begin{figure}[ht]
\hspace{0cm}
\includegraphics[scale=0.55]{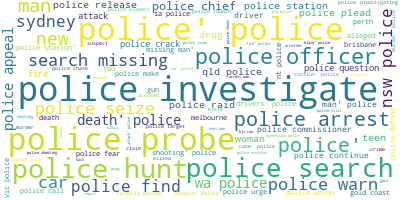}
\caption{Nuvem de palavras usando as \textit{headlines} do cluster $15$, considerando $k=19$ clusters e com 1-\textit{gram} + 2-\textit{gram}. \label{fig9}}   
\end{figure}

\begin{figure}[ht]
\hspace{0cm}
\includegraphics[scale=0.55]{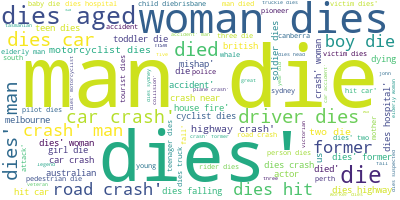}
\caption{Nuvem de palavras usando as \textit{headlines} do cluster $19$, considerando $k=19$ clusters e com 1-\textit{gram} + 2-\textit{gram}. \label{fig10}}   
\end{figure}

 A Tabela \ref{Tab5} apresenta os principais temas dos cluster criados pelo método \textit{K-means} considerando $k = 19$ para o caso  1-\textit{gram} + 2-\textit{gram}. As Figuras \ref{fig9} e \ref{fig10} mostram as nuvens de palavras usando as \textit{headlines} dos clusters referentes a "\textit{police}"  \ e "\textit{dies}".

\begin{table}[ht]
\centering
\scalebox{0.7}{ 
\begin{tabular}{|l|c|c|c|c|c|c|c|c|c|c|c|} 
\hline
\rowcolor{cyan}
Cluster &  1 & 2 & 3 & 4 & 5 \\
Tema & * & us & say & * & prices  \\
\hline
\hline
\rowcolor{cyan}
Cluster &  6 & 7 & 8 & 9 & 10 \\
Tema & hospital & mayor & Hewitt & New York & New Zeland  \\
\hline
\hline
\rowcolor{cyan}
Cluster &  11 & 12 & 13 & 14 & 15 \\ 
Tema &  hit-run & missing & face & court & police \\
\hline
\hline
\rowcolor{cyan}
Cluster & 16 & 17 & 18 & 19 & - \\
Tema & * & trial & council & dies & - \\
\hline
\end{tabular}}
\caption{Principais palavras dos clusters considerando $19$ clusters e com 1-\textit{gram} e 2-\textit{gram}. * palavras variadas.} \label{Tab5}
\end{table}

\subsection{Clusters para cada ano}

\begin{figure}[ht]
\hspace{-1.2cm}
\includegraphics[scale=0.235]{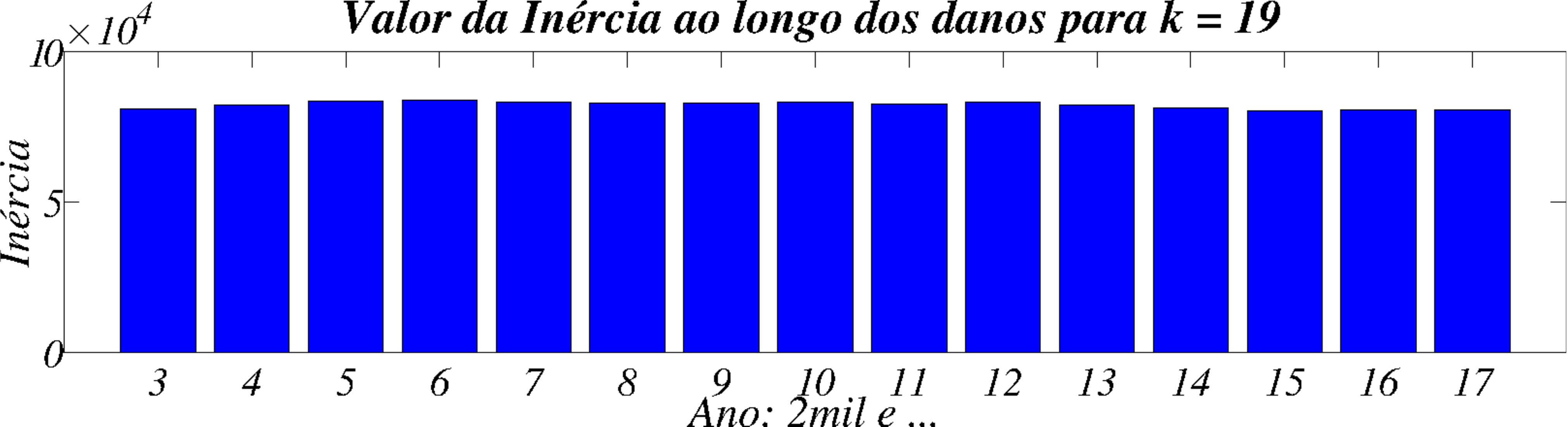}
\caption{Valores da função inércia \ref{eq1} em relação aos anos e considerando $k=19$ clusters e 1-\textit{gram} + 2-\textit{gram}. \label{fig4}}   
\end{figure}

\begin{figure}[ht]
\hspace{-3.5cm}
\includegraphics[scale=0.3]{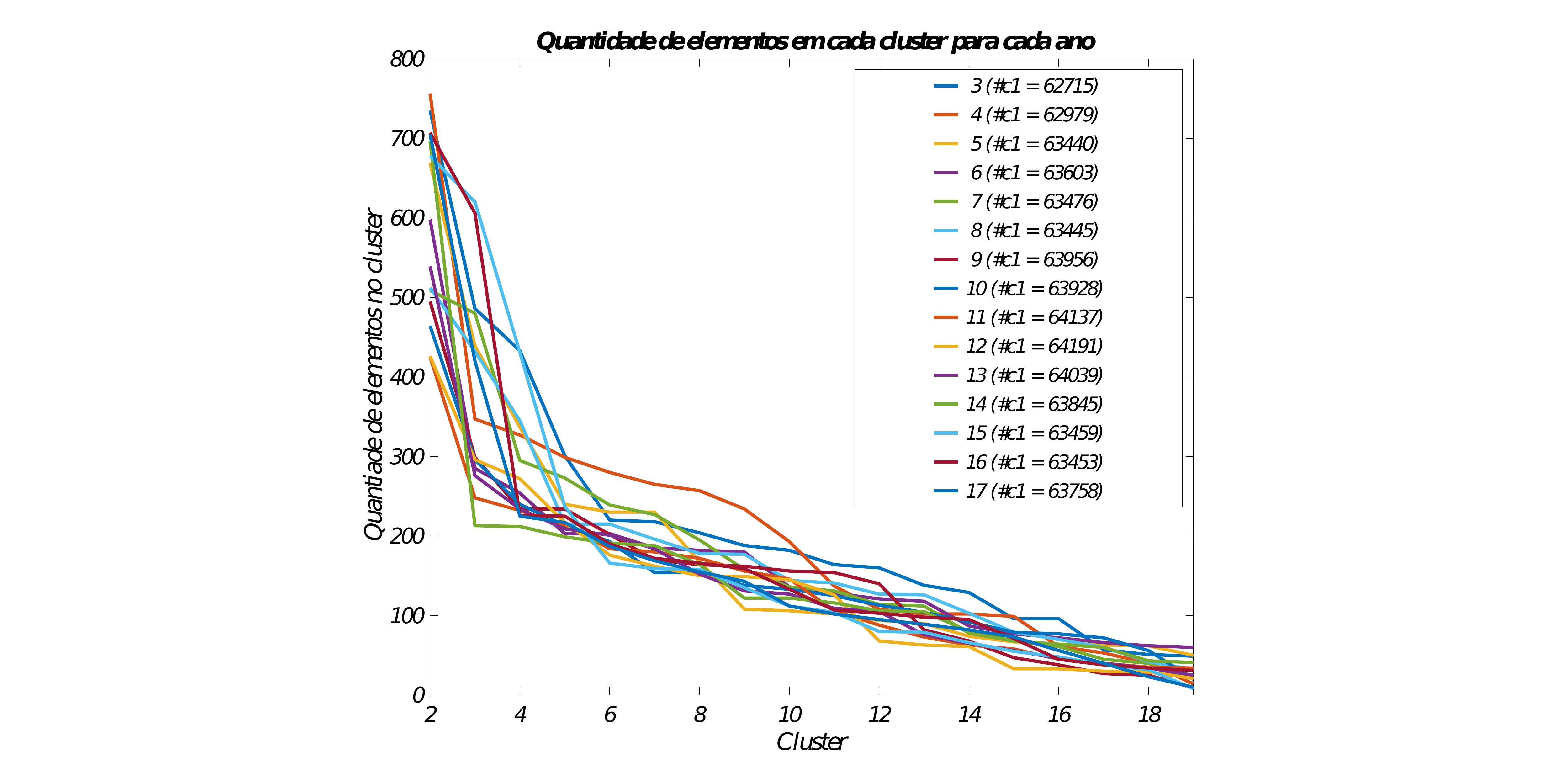}
\caption{Cardinalidade de cada \textit{cluster} para cada anos e considerando $k=19$ clusters e 1-\textit{gram} + 2-\textit{gram}. \label{fig5}}   
\end{figure}

Usando $k=19$ dividimos o conjunto de \textit{headlines} por anos ($15$ subconjuntos no total), e aplicamos \textit{k-means} em cada um deles. A Figura \ref{fig4} apresenta os valores da inércia para cada ano (2013-2017), nela podemos ver que a variação para cada ano é bem pequena. Na Figura \ref{fig5} podemos observar a quantidade de elementos em cada \textit{cluster} em cada ano, para todos os anos a cardinalidade do \textit{cluster} com maior quantidade elementos foi representado na legenda.


\section{Conclusões e Trabalhos Futuros}

Os resultados indicam que a utilização de \textit{k-means} separa os dados devagar tendo sempre um cluster maior que vai diminuindo de tamanho na medida que aumentamos $k$. Achamos que isto acontece devido ao fato de que as \textit{headlines} são cadeias de texto muito pequenas, entre $8$ e $10$ palavras, e temas que deveriam ser agrupados não conseguem se conectar, por exemplo, se o tema for esportes poderia ter \textit{headlines} falando de tênis e futebol usando palavras diferentes. 

Podemos observar consistência nos resultados. Por um lado a Figura \ref{fig4} mostra que os valores da função inércia são similares para todos os anos e por outro, a Figura \ref{fig5} indica que a quantidade de elementos de cada clusters ano a ano se é similar o qual seria esperado se o agrupamento estivesse relacionando os mesmos temas.

O uso de \textit{3-grams} e \textit{4-grams} pode melhorar o valor da inércia, estes atributos foram calculados porém os códigos demoravam  para realizar o agrupamento e por isso ficamos apenas no caso $N=1,2$. 
  
Além disso as notícias costumam ser muito localizadas no tempo, assim as mais conectadas seriam as referentes a um mesmo acontecimento (por exemplo "\textit{brazil world cup}") que formariam um cluster pequeno de algumas dezenas ou centenas de \textit{headlines}. Isto explicaria o alto valor da função inércia. Valores menores da inércia seriam alcançados só quando o valor de $k$ é muito grande. Provavelmente um resumo da notícia (chamadas) permitiria que aparecessem nos dados mais palavras associadas ao tema que os relacionam e teríamos um melhor agrupamento.







\end{document}